\documentstyle[leqno]{bcp01e}
\input amssym.tex
\input amssym.def

\font\tendl=msbm10  % scaled \magstep1   % double line
\font\sevendl=msbm7 % scaled \magstep1
\font\fivedl=msbm5  % scaled \magstep1
\font\tengl=eufm10  % scaled \magstep1   % gothic letters
\font\sevengl=eufm7 % scaled \magstep1
\font\fivegl=eufm5  % scaled \magstep1

\newfam\dlfam \def\dl{\fam\dlfam\tendl} % \dl is double line
\textfont\dlfam=\tendl \scriptfont\dlfam=\sevendl
\scriptscriptfont\dlfam=\fivedl
\newfam\glfam \def\gl{\fam\glfam\tengl} % \gl is gothic letters
\textfont\glfam=\tengl \scriptfont\glfam=\sevengl
\scriptscriptfont\glfam=\fivegl

\newcommand{\CE}{{\cal E}}

\newcommand{\CH}{{\cal H}}

\newcommand{\CJ}{{\cal J}}

\newcommand{\CL}{{\cal L}}

\newcommand{\CO}{{\cal O}}

\newcommand{\CR}{{\cal R}}
\newcommand{\CS}{{\cal S}}

\newcommand{\CW}{{\cal W}}

\newcommand{\CX}{{\cal X}}

\def\mathbb#1{\dl #1}

\newcommand{\NR}{{{\mathbb R}}}

\newcommand{\NC}{{{\mathbb C}}}

\newcommand{\NZ}{{{\mathbb Z}}}

\newcommand{\NN}{{{\mathbb N}}}

\mathchardef\zx="7118  %\xi
\mathchardef\zl="7115 

\newtheorem{lemma}{Lemma}
\newtheorem{coro}{Corollary}
\newtheorem{theo}{Proposition}
\newtheorem{Theo}{Theorem}

\newcommand{\be}{\begin{equation}}
\newcommand{\ee}{\end{equation}}

\catcode `\@=11
\@addtoreset{equation}{section}
\catcode `\@=12

\newcommand{\joint}[1]{\mathop{{\rm ad}(#1)}\nolimits}
\newcommand{\pol}{\frac{1}{2}}

\newcommand{\AB}{\allowbreak}
\newcommand{\gr}[2]{\mathop{{\bf #1}(#2)}\nolimits}

\newcommand{\BDS}{\mathop{\ts\bigoplus}\limits}

\newcommand{\ad}{\mathop{\rm ad}\nolimits}
\newcommand{\ADA}[1]{\ifmmode \ad(#1) \else $\ad(#1)$\fi}
\newcommand{\LI}[2]{\ifmmode#2_1,\AB\,\ldots,\,\AB #2_{#1}%
\else$ #2_1,\AB\,\ldots,\,\AB#2_{#1}$\fi}

\newcommand{\fr}{\gl}

\newcommand{\ts}{\textstyle}

\newcommand{\eq}{\hspace*{-.5em} = \hspace*{-.5em}}

\def\longlongrightarrow{%
\relbar\joinrel\relbar\joinrel\longrightarrow}

\begin{document}

%%%%% FOR ARTICLE.cls %%%%%%%

%\title{%\vspace*{-1cm}
%The Weyl Algebra, Spherical Harmonics,\\
%and Hahn Polynomials}

%\author{\bf Ewa\ Gnatowska\\ %}\address{
%\footnotesize{Department of Mathematical Methods of Physics,}\\[-6pt]
%\footnotesize{Faculty of Physics,\ University of Warsaw,}\\[-6pt]
%\footnotesize{Ho\.za 74, 00-682 Warsaw, Poland}\\[-6pt]
%\footnotesize{E-mail: gnatowsk@fuw.edu.pl}
%\and 
%and\\
%\author{
%\bf Aleksander\ Strasburger\\%}\address{
%\footnotesize{Institute of Mathematics, University of Bia\l{}ystok,}\\[-6pt]
%\footnotesize{Akademicka 2, 15-267 Bia\l{}ystok}\\[-6pt]
%\footnotesize{\it and}\\[-6pt]
%\footnotesize{Department of Mathematical Methods of Physics,}\\[-6pt]
%\footnotesize{Faculty of Physics,\ University of Warsaw,}\\[-6pt]
%\footnotesize{Ho\.za 74, 00-682 Warsaw, Poland}\\[-6pt]
%\footnotesize{E-mail: alekstra@math.uwb.edu.pl}
%}

%\date{}

%\maketitle

%%%%% END  ARTICLE.STY %%%%%%%

%%%%% FOR BCP01e.STY %%%%%%%

\newbox\keybox
\newcount\knum
\long\def\keywords#1{\global\knum=100
\setbox\keybox=\hbox{\vfootnote{}{{\it Key words
and phrases}\/: #1}{}}}
\title{%
The Weyl Algebra, Spherical Harmonics,\\
and Hahn Polynomials}

\author{Ewa\ Gnatowska}
\address{Department of Mathematical Methods of Physics,
Faculty of Physics,\ University of Warsaw\\
Ho\.za 74, 00-682 Warsaw, Poland\\
E-mail: gnatowsk@fuw.edu.pl}
\author{Aleksander\ Strasburger}
\address{Institute of Mathematics, University of Bia\l{}ystok\\
Akademicka 2, 15-267 Bia\l{}ystok\\
{\it and}\\
Department of Mathematical Methods of Physics,
Faculty of Physics,\ University of Warsaw\\
Ho\.za 74, 00-682 Warsaw, Poland\\
E-mail: alekstra@math.uwb.edu.pl}
\mathclass{Primary 81R10, 33C55; Secondary: 16W25, 33C45, 33C80.}

\maketitlebcp

%%%%% END BCP01E.STY %%%%%%%

%\begin{abstract} 
%\noindent 
\abstract{In this article we apply the duality technique of R. Howe to study 
the structure of the Weyl algebra. We introduce a one-parameter family of 
``ordering maps'', where by an ordering map  we understand a vector space isomorphism of the 
polynomial algebra on $\NR^{2d}$ with the Weyl algebra generated by 
creation and annihilation operators $a_1,\,\ldots,\,a_d,\,a_1^+,\,\ldots,\,a_d^+$.
Corresponding to these orderings, we construct a one-parameter family of 
${\fr sl}_2$ actions on the Weyl algebra, what enables us to 
define and study  certain subspaces of the Weyl algebra 
--- the space of Weyl spherical harmonics and the space of ``radial polynomials''.
For the latter we generalize results of Luck and Biedenharn, 
Bender et al., and Koornwinder describing the radial elements in terms of 
continuous Hahn polynomials of the number operator.
}

%\vspace{0.2cm} 
%\footnotesize{\noindent 2000 {\it Mathematics Subject Classification}\/: 
%Primary 81R10, 33C55; Secondary: 16W25, 33C45, 33C80.} 
%\vspace{.2cm} 
%\noindent{\bf Key words:} {\em  Weyl algebra, canonical commutation relations,
%ordering map, Howe's duality, ${\fr sl}_2$ weight module, harmonic polynomials, orthogonal polynomials, hypergeometric 
%function.} 
%\end{abstract} 

\section{Introduction.}

This article reports on an investigation (still in progress) 
of the Weyl algebra with the aid of  R. Howe's duality technique.
More specifically, we attempt to transfer the
information obtained by considering the action on the (commutative)
polynomial algebra on $\NR^{2d}$ of a triple of classical
operators (consisting of the Laplacian, Euler operator and multiplication
by the square of the radius and giving rise to an ${\fr sl}_2$ action
on this algebra)  to the case of the (noncommutative) Weyl algebra
with $2d$ generators --- see the beginning of Section 3 for the
definition of the Weyl algebra.  On this route one is inevitably
confronted with the renowned ``ordering problem of quantum mechanics'',
c.f. e.g.  \cite{KBW,Be2,BD}. We deal with this problem in a wider context
by introducing a one-parameter family of ``orderings'', by which we
understand a family of vector space isomorphisms of the polynomial algebra
with the Weyl algebra. Corresponding to these orderings, we construct a
one-parameter family of ${\fr sl}_2$ representations acting on the Weyl algebra,
whose action we employ to decompose the algebra.

Two subspaces of the Weyl algebra stand out in our analysis
--- the subspace of ``Weyl spherical harmonics'' and the (in a certain sense
complementary) subspace of ``radial'' polynomials, which are polynomials
of a single element, the so-called number operator.  In this paper
we concentrate on the latter one, emphasizing the connection with the
problems of the special function theory.  This is the subject of the
second part of the paper, where we describe properties of these radial
polynomials in various orderings. In particular we identify
the elements of a natural basis of the space with Hahn polynomials of the
shifted number operator.  This is a generalization of results obtained
earlier by Lohe, Biedenharn and Louck \cite{LB}, Bender, Mead, and
Pinsky \cite{Be2}, and Koornwinder \cite{KO}, among others.

\section{Preliminaries on the polynomial algebra in even dimension.}

Consider the real cartesian space of even dimension
$\NR^{2d}$, $d\ge1$, with coordinates denoted by $\LI{d}{x},\,\LI{d}{\zx}$.
Let $P=P_\NC(\NR^{2d})$ denote the polynomial algebra with complex
coefficients over $\NR^{2d}$ and $P^k=P^k_\NC(\NR^{2d})$ its subspace
consisting of homogeneous polynomials of degree $k$. Setting $z_j=x_j+i\zx_j$
and   $\bar{z_j}=x_j-i\zx_j$ for $j=1,\,\ldots,\,d$, where here and
everywhere in  the sequel $i=\sqrt{-1}$, we can regard  $P$ as
the polynomial algebra with respect to coordinate functions $ z_1,\, \ldots,\,
z_d$ and their conjugates $ \bar{z_1},\, \ldots,\, \bar{z_d}$.  Thus, using
the usual multi-index notation we shall write its elements in the form
\begin{eqnarray}
 \label{Pe}
p(z, \bar{z}) = \sum_{ \alpha , \beta} p_{\alpha ,\beta}  z^{\alpha }
\bar{z}^{\beta },
\end{eqnarray}
where $z=(z_1,\,\ldots,\,z_d)$,
$\bar{z}=(\bar{z}_1,\,\ldots,\,\bar{z}_d)\in \NC^d$ and $ p_{\alpha ,\beta}\in \NC$.
In particular,
$$r^2 = \sum_{j=1}^{d}x_j^2 + \sum_{j=1}^{d} \zx_j^2=
\sum_{j=1}^{d} z_j \bar{z}_j
$$
is the square-of-the-radius function (euclidean length squared) on $\NR^{2d}$.

Similarly, setting as usual
$$ \frac{\partial }{\partial z_j} = \frac{1}{2}
\biggl( \frac{\partial }{\partial x_j}
- i \frac{\partial }{\partial \zx_j}\biggr ), \qquad
\frac{\partial }{\partial \bar{z}_j} = \frac{1}{2}
\biggl( \frac{\partial }{\partial x_j}
+ i \frac{\partial }{\partial \zx_j}\biggr),
$$
one can express in the complex form the euclidean Laplace operator (Laplacian)
$$
\Delta = \sum_{j=1}^{d}\frac{ \partial^2}{\partial x_j^2}
+ \sum_{j=1}^{d} \frac{\partial^2}
{\partial \zx_j^2} = 4 \sum_{j=1}^{d} \frac{\partial^2}{\partial z_j
\partial \bar{z}_j },
$$
and the Euler operator
$$
\widetilde{E}  = \sum_{j=1}^{d} x_j \frac{\partial }{\partial x_j} + \sum_{j=1}^{d}
\zx_j \frac{\partial }{\partial \zx_j} =  \sum_{j=1}^{d}
 z_j \frac{\partial }{\partial z_j} + \sum_{j=1}^{d}\bar{z_j}
\frac{ \partial }{\partial \bar{z_j}}.
$$

Along with the usual action of the orthogonal group $\gr{O}{2d}$
on $P$, we shall be concerned with the action of the unitary
group $\gr{U}{d}$ on $P$ which is obtained by
extending the action on generators given by
\begin{eqnarray*}
z\mapsto gz,\qquad \bar{z}\mapsto
\bar{g}\bar{z},\qquad g\in \gr{U}{d},
\end{eqnarray*}
where $z=(z_1,\,\ldots,\,z_d)$ and $\bar{z}=(\bar{z}_1,\,\ldots,\,\bar{z}_d)$
are regarded as column vectors. Recall that the contragredient matrix
$(g^t)^{-1}$ of any $g\in \gr{U}{d}$ satisfies $(g^t)^{-1} =\bar{g}$, the
bar denoting complex conjugation (taken entry-wise). These two actions are
related by the natural embedding of $\gr{U}{d}$ into $\gr{O}{2d}$.

Modifying slightly the above differential operators and 
supplementing them by the multiplication by the square of the
radius one obtains a triple of endomorphisms inducing an 
action of the Lie algebra ${\fr sl}_2$ on $P$. In fact, it can be 
easily checked that by setting 
\begin{eqnarray}
\label{rle}
R p =r^2 p, \qquad
L p  = \frac{1}{4} \Delta p, \qquad
E p = \bigl(\widetilde{E} + d \bigr)p , \qquad \mbox{for\ }p\in P,
\end{eqnarray}
(here $\widetilde{E} + d $ is the symmetrized Euler operator)
we obtain commutation rules for
the standard generators of the Lie algebra $ sl_2 $, i.e.
\begin{equation}
\label{sl2}
[R,\, L] = -E
\qquad
[E, \,R] = 2 R
\qquad
[E, \,L] = - 2 L. 
\end{equation}

It can be verified that this latter action commutes with the 
action of $\gr{O}{2d}$. 
A remarkable although elementary application of Howe's duality
principle based on using the above actions 
is the analysis of the structure of the algebra $P$
(cf. \cite{HO}, p.~118), which gives a large part of the
classical theory of spherical harmonics.
In particular one arrives this way at the decomposition
$P = J \otimes H$, where $J\subset P$ is the subalgebra of
$\gr{O}{2d}$ invariant polynomials, known to be the subalgebra 
generated by $r^{2}$, and $H = H(\NR^{2d})= \{p \in P \mid \Delta p = 0 \}$ is the
subspace of harmonic polynomials in $P$. The isomorphism is
obtained via multiplication. Consequently each homogeneous
polynomial $ p \in P$  can be decomposed into a sum of products
of homogeneous harmonic polynomials with powers of the 
radius-square; 
\begin{eqnarray*} p (z, \bar{z}) = \sum_{j=0}^{[
\frac{1}{2} \deg p ]} r^{2j} h_j (z,\, \bar{z}),
\end{eqnarray*} where $h_j$ are homogeneous harmonic
polynomials  of degree $\deg p - 2j$. (See e.g. \cite{FA} for a
derivation of this decomposition along classical lines).

Furthermore, the following facts can be obtained within this
approach.

Let $H^k \subset H$ denote the subspace of
homogeneous harmonic polynomials of degree $k$.

a) For each  $k\in\NN$ the subspace $H^k$ is irreducible under the
natural action of $\gr{O}{2d}$ and for any $j\neq k$ the actions on $H^j$ and $H^k$ are  not equivalent. Therefore the  direct sum decomposition
$H = \oplus_{j=0}^{\infty} H^j $ is multiplicity free.

b) For each $m\in \NN$ the space  $P^m$ decomposes into a direct
sum of $\gr{O}{2d}$ irreducible subspaces as
\begin{eqnarray}
\label{roz}
P^m=\BDS_{j=0}^{[\frac{1}{2}m]} r^{2j}H^{m-2j}.
\end{eqnarray}

c) Let $P(\NR^{2d};H^k)$ be the sum of all $\gr{O}{2d}$ irreducible
subspaces of $P$ equivalent to $H^k$; it is called isotypic subspace of
type $H^k$. Then the multiplication
\begin{eqnarray}
\label{ten}
J \otimes H^k \to P(\NR^{2d};H^k);\qquad  
p(r^2)\otimes h(z,\,\bar{z})\mapsto p(r^2)\cdot h(z,\,\bar{z})
\end{eqnarray}
induces an equivariant isomorphism of $P(\NR^{2d};H^k)$ with $J \otimes H^k$.

As the actions of  ${\fr sl}_2$ and $\gr{O}{2d}$ commute with each
other, it makes sense to consider their joint action --- perhaps 
more properly one should speak of the action of the direct product 
of the Lie algebras ${\fr sl}_2$ and $o(2d)$ on $P$. It can be shown that
the space $P(\NR^{2d};H^k)$ is irreducible under this joint action.
By transfering the action of ${\fr sl}_2$ to the algebra of
invariants $J=P[r^2]$ by means of the multiplication map 
(\ref{ten}) one obtains the action which is (algebraically) equivalent to that of
one of the lowest weight representations of ${\fr sl}_2$ --- the lowest weight
(with respect to $E$) being equal to $(k+d)$. Actually, the lowest
weight vectors in this space can be identified as certain orthogonal
polynomials. 
%(see e.g. \cite{HO}\,[Chapter III, p. 118]).

Lastly, it is known, that under restriction to $\gr{U}{d}\subset
\gr{O}{2d}$  the irreducible $\gr{O}{2d}$-space $H^k $ splits into the
direct sum of $\gr{U}{d}$-irreducible subspaces,
\begin{equation} \label{bi-deg_decomp}
H^k = \BDS_{m+n=k}H^{(n,m)}
\end{equation}
where $H^{(n,m)}$ denotes the space of
homogeneous polynomials of bi-degree $(n,\,m)$. We recall that
$p(z,\,\bar{z})\in H^{(n,m)} \iff p(\zl z,\,\bar{\zl}\bar{z})=
\zl^n\bar{\zl}^m p(z,\,\bar{z})$. 
Consequently, the isotypic subspaces of type $H^{(n,m)}$ in $P$   
are of the form $J\cdot H^{(n,m)}$, and 
$$
P(\NR^{2d};H^k) = \BDS_{m+n=k}J\cdot H^{(n,m)}.
$$
For the explicit procedure of
obtaining the decomposition (\ref{bi-deg_decomp}), in
particular for the formulae expressing projections onto subspaces of the
given bi-degree, we refer to \cite{IK}.

\section{Construction of an ${\fr sl}_2$ action on the Weyl algebra.}

In this section we shall describe a method of transfering the results
concerning the polynomial algebra described at the end of the preceeding
section to the case of the Weyl algebra.

Usually, by the Weyl algebra with $2d$ generators $\CW =\CW_{2d}$ 
one means the associative algebra with unit $I$ generated by $2d$ 
elements $a_1,\,\ldots,\,a_d,\,a_1^+,\,\ldots,\,a_d^+$ subject 
to the relations (called Canonical Commutation Relations --- 
abbreviated CCR in the sequel)
\begin{equation}
\label{CCR}
\qquad a_j a_k^+ - a_k^+ a_j =  \delta_{kj} I,\quad
a_j a_k - a_k a_j =0,\quad  a_k^+ a_j^+ - a_k^+ a_j^+=0,\quad
1 \leq j, k \leq d.
\end{equation}

We shall employ the following standard notation from
the theory of associative algebras: for arbitrary
$x,\,y\in\CW$  we shall denote by $[x,\,y]=xy-yx $
the commutator of $x$ and $y$ and by $\{ x,y \} = xy +yx$ 
their anticommutator. For $x \in\CW$ the adjoint map 
$\joint{x}:\CW\to \CW $ is defined by setting 
$\joint{x}w=[x,\,w]$ for  $w\in \CW$ --- as well-known,
$\joint{x}$ is a derivation of $\CW$.

Let us recall that for many questions it is more convenient to
realize the abstractly defined Weyl algebra as the algebra of partial
differential operators with polynomial coefficients, considered initially
as operators acting as on the Schwartz space $\CS(\NR^d)$, and regarded,
when need arrises, as densely defined operators in $L^2(\NR^d,d\lambda(x))$
--- the space of square integrable functions on $\NR^d$
with respect to the Lebesgue measure. This identification is obtained  by the
standard assignment
\[
 a_{j}=\frac{1}{\sqrt2}\biggl(x_j+\frac{\partial}{\partial
x_j}\biggr),  \qquad 
a_j^+=\frac{1}{\sqrt2}\biggl(x_j-\frac{\partial}{\partial
x_j}\biggr).
\]
Note that $a_j^*=a_j^+$, where the adjoint ${}^*$
operation is understood in the sense of the formal adjoint of a differential
operator. In quantum physics the operators $a_j$, $a_j^+$ are known as
annihilation and creation operators, respectively. 
For the present work however, it is the algebraic structure of 
the Weyl algebra $\CW$  that matters. 
 
Given  $a \in \CW$ we  denote  by $L_a$ the map $\CW \to \CW$ obtained by the
left multiplication with $a$ and by $L_a^+$ the map corresponding to the left
multiplication by $a^+$. The symbols $R_a$, $R_a^+$ will have analogous
meanings with respect to the right multiplication. By taking convex combinations
of left and right multiplication operators we shall be
able to construct a family of endomorphisms of $\CW$ which will serve as a
replacement for the operators of multiplication by coordinates and their
complex conjugates in the polynomial algebra.

\defin{Definition 1.}{For a given $a\in \CW$ and $q \in [0,1]$ we
define(\footnote{The condition $q \in [0,1]$ turns out to be nonessential and
will be removed in further publications.}) endomorphisms $M_a$ 
and $M_a^+$ of $\CW$ by the formulae}
\begin{eqnarray} \label{emy}
M_a   & \eq &(1-q)L_a + q R_a = L_a - q \joint{a} \\
M_a^+ & \eq & q L_a^+ + (1-q) R_a^+ = L_a^+ - (1-q) \joint{a^+}.
\end{eqnarray}

Let us explicitly point out to the meaning of this definition
for the special values $q=0,\pol,1$ of the parameter. For $q=0$, $M_a$
is the left multiplication by $a$, while $M_a^+$ is the right multiplication
by $a^+$. For $q=1$ these roles are reversed, while for $q=\pol$ each is
equal to $1/2$ of the respective anticommutator.
\begin{lemma}
Let $\{a_1,\,\ldots,\,a_d,\,a_1^+,\,\ldots,\,a_d^+\}$ be a  set of generators of
the Weyl algebra and put  $M_j = M_{a_j},\quad M_j^+ = M^+_{a_j}$,
$j=1,\ldots,d$.  Then $\{ M_1,\,\ldots,\,M_d,\,M_1^+,\,\allowbreak \ldots,\,M_d^+\}$ is a
commutative  family of endomorphisms of $\CW$.
\end{lemma}
\Proof In view of CCR it is clear that the $M_j$'s ($M_j^+$'s respectively)
commute between themselves, and that $[M_j,M^+_k] = 0$ for $j \not = k$.
This last relation is also true for $j=k$. In fact
\[
[M_j, M_j^{+}] = (1-q)q ( [L_{a_j},L_{a_j}^+] +[R_{a_j} ,R_{a_j}^+] )
= (1-q)q (I-I) = 0.\hfill \Box
\]

Thus employing the multi-index notation for the family $\{ M_1,\ldots,
M_d,M_1^+ ,\allowbreak \ldots,M_d^+ \}$, i.e.  writing $M^{\alpha} M^{+\beta} =
M^{\alpha_1}_1 \ldots M^{\alpha_d}_d M^{+\beta_1}_1 \ldots
M^{+\beta_d}_d$ for $\alpha, \beta \in \NZ^d_+$, we have by virtue
of the lemma
\begin{equation}
M^{\alpha} M^{+\beta}\cdot M^{\alpha'} M^{+\beta'}=
M^{(\alpha+\alpha')} M^{+(\beta+\beta')}.
\end{equation}

\begin{lemma}
The map
$\CO_q : P(\NR^{2d}) \to \CW_{2d}$ defined by
\begin{eqnarray}
\label{MM}
\CO_q (z^{\alpha} {\bar z}^{\beta}) = M^{\alpha} M^{+ \beta} I
\end{eqnarray}
is a vector space isomorphism such that
\begin{eqnarray}
\CO_q(z_jp) = M_j \CO_q(p), &&
\CO_q(\bar{z_j}p) = M^+_j \CO_q(p),
\label{begin_first_step} \\
\CO_q\biggl(\frac{\partial p}{\partial z_j}\biggr) =
- \joint{ a_j^+} \CO_q(p),&&
\CO_q\biggl(\frac{\partial p}{\partial \bar{z_j}}\biggr) = 
\joint{ a_j} \CO_q(p).
\label{begin_second_step}
\end{eqnarray}
\end{lemma}
\Proof
The first pair of relations is obvious by the definition of $\CO_q$ and 
Lemma 1.  The second pair, concerning derivations, follows by routine 
calculation. \hfill $\Box$

\defin{Definition 2.}{The map $\CO_q :P (\NR^{2d}) \to \CW_{2d}$ 
defined for $q \in [0,1]$  by the equation (\ref{MM}) will be called 
the $q$-ordering map.} 

Since the subalgebras of $\CW$ generated by pairs $\{ a_l,a_l^+\}$,
$\{a_k,a_k^+\}$ with $k \not = l$, commute with each other, the elements
(\ref{MM}) can be written in the factorized form,
 \begin{eqnarray*}
\CO_q (z^{\alpha} {\bar z}^{\beta}) = \prod_{l=1}^d B^{\alpha_l}_{\beta_l}
(a_l,a_l^+),
\end{eqnarray*}
where for nonnegative integers $k,j$ we have set
\begin{eqnarray}
\label{Bs}
B^{j}_{k} (a_l,a_l^+) = \sum_{s=0}^{k} {k \choose s}
q^{k - s} (1-q)^{s} a_l^{s} (a_l^+)^{j} a_l^{k - s}.
\end{eqnarray}

In the special cases mentioned above the formula (\ref{Bs}) specialises to the
following well-known constructions.

a) For $q=0$ it gives the so-called normal (Wick) ordering
\[
\CO_0(z^{\alpha} {\bar z}^{\beta}) = a^{\alpha} (a^+)^{\beta},
\]
with all $a_j$ (annihilation operators) on the left and all $a_j^+$ (creation
operators) on the right.

b) For $q=1$ the roles of $a_j$ and $a_j^+$ are reversed and $\CO_1$ gives the
anti-normal (anti-Wick) ordering
\[
\CO_1(z^{\alpha} {\bar z}^{\beta}) = a^{+\beta} a^{\alpha}.
\]

c) The case $q= \pol$ corresponds to the so-called Weyl, or symmetric,
ordering.
\medskip

All these assertions are evident on the basis of (\ref{emy}).
Further, let us observe that our elements $B^{k}_{j}$ are related
to the basis elements $P^j_m$ introduced for the case of one degree of freedom
($d=1$) by Louck and Biedenharn in \cite{BL}. In fact, $B^{k}_{j}$ is
proportional to
\[
 P^{\frac{j-k}{2}}_{\frac{j+k}{2}} (a,a^+) =
2^{\frac{j-k}{2}}\sqrt{\frac{k!}{j!}} \sum_{l=0}^{k} \frac{a^l (a^+)^j
a^{k-l}}{l!(j-l)!}.
\]
Since the $q$-ordering map $\CO_q $ is for each value of
$q$ an isomorphism of vector spaces, there exist endomorphisms 
of the Weyl algebra, which we shall denote $\CR_q,\ \CL_q$, and 
$\CE_q$ (for simplicity we suppress the index $q$ writing 
$\CR=\CR_q$ 
etc., when no confusion can arise), such that the following 
equalities hold
\begin{eqnarray} \label{intertwining}
 \qquad \CO_q R= \CR_q \CO_q ,\qquad \CO_q L= \CL_q \CO_q ,
 \qquad \CO_q E = \CE_q \CO_q. 
\end{eqnarray}
This can be conveniently summarised in the form of 
commutative diagrams
\begin{equation} \label{diag}
\matrix{
P_\NC(\NR^{2d}) & {{R ,\, L,\, E }\atop \longlongrightarrow }
&       P_\NC(\NR^{2d}) \cr
 \CO_q \Bigg\downarrow & &  \CO_q\Bigg\downarrow \cr
 \CW_{2d} & {{\CR_q ,\, \CL_q,\, \CE_q }\atop \longlongrightarrow } &
 \CW_{2d}}
\end{equation}

>From (\ref{rle}) and 
(\ref{begin_first_step}--\ref{begin_second_step})
we can deduce explicit expressions for 
$\CR_q,\,\CL_q$ and $\CE_q$.

\begin{lemma} \label{triple}
The endomorphisms $\CR_q,\,\CL_q,\,\CE_q$ of the Weyl algebra 
$\CW_{2d}$
are given by the following formulae:
\begin{eqnarray}
\label{*}
\qquad \qquad \CR_q w
\hspace*{-.5em} &=& \hspace*{-.5em}
(1-q)^2\sum_{j=1}^{d}a_j w a_j^+ + q(1-q)
\sum_{j=1}^{d} ( w a_j^+a_j + a_ja_j^+ w ) + 
q^2 \sum_{j=1}^{d} a_j^+ wa_j,\\
\qquad\qquad \CL_q  w
\hspace*{-.5em} &=& \hspace*{-.5em}
 - \sum_{j=1}^{d} \joint{ a_j} \joint{a_j^+} w , \label{L_q}\\
\qquad\qquad \CE_q w
\hspace*{-.5em} &=& \hspace*{-.5em}
-\sum_{j=1}^{d}\{(1-q)(a_j[a_j^+, w]-[a_j, w]a_j^+) +q([a_j^+,
w]a_j-a_j^+[a_j, w])\} + d w .
 \end{eqnarray}
\end{lemma}
Note that $\CL_q$ does not depend on $q$ and moreover, since 
$\joint{a_j} \joint{a_j^+} = \joint{a_j^+} \joint{a_j}$, 
we also have $\CL_q = - \sum_{j=1}^d \joint{a_j^+} \joint{a_j} $.

For the case of the Weyl (symmetric) ordering, i.e. 
for $q=\pol$, we shall employ the notation $\CR_s$, $\CL_s$, 
$\CE_s$ 
for these operators. Note that they are given by fairly simple
expressions (recall $\{a,b\}$ stands for the anticommutator of $a,\,b\in \CW$):
\begin{eqnarray*}
\CR_s w &\eq & \frac{1}{4} \sum_{j=1}^d \{ a_j, \{ a_j^+, 
w \} \}, \\
\CL_s  w &\eq &- \sum_{j=1}^d [ a_j, [ a_j^+, w ] ], \\
\CE_s  w &\eq & \sum_{j=1}^d ( a_j  w  a_j^+
-  a_j^+ w a_j  ).
\end{eqnarray*}
% Writing the commutators and anti-commutators
%explicitly we get equivalent form
%%\begin{eqnarray*}
%\CR_s  w &=& \frac{1}{4} \sum_{j=1}^d
%( a_j^+ a_j  w + a_j  w   a_j^+
%+ a_j^+  w a_j +  w a_j a_j^+ ), \\
%\CL_s w &=&  \sum_{j=1}^d (- a_j^+ a_j w + a_j w  a_j^+
%+ a_j^+ w a_j - w a_j a_j^+ ), \\
%\CE_s w &=& \sum_{j=1}^d ( a_j w  a_j^+
%- a_j^+  w a_j  ).
%\end{eqnarray*}
Applying successively the defining relation (\ref{intertwining}) to the 
commutation relations (\ref{sl2}), we see that the endomorphisms
$\CR_q$, $\CL_q$ and $\CE_q$ of the Weyl algebra 
also satisfy the commutation relations of the ${\fr sl}_2$ Lie algebra, 
namely
\begin{eqnarray}
\label{SL2}
 [\CR_q, \CL_q] = - \CE_q, \qquad [\CE_q, \CR_q] = 2 \CR_q, 
 \qquad [\CE_q, \CL_q] = - 2 \CL_q
\end{eqnarray}
(which can also be checked by brute force calculation).
This leads us to the following important result. 
\begin{coro} \label{wtriple}
The $q$-ordering map $ \CO_q : P (\NR^{2d}) \to \CW_{2d} $ is an
intertwining map  for actions of the Lie algebra  ${\fr sl}_2$ realized by the
triples $\{R ,\,L, \,E \}$ and $\{\CR_q,\,\CL_q,\,\CE_q \}$. 
\end{coro}

\section{The analysis of radial Weyl polynomials.}

We now apply the idea of R. Howe to use the action of ${\fr sl}_2$ 
determined by $\{\CR_q,\,\CL_q,\,\CE_q \}$ to decompose the Weyl algebra $\CW$.
Given any homogenous polynomial $p \in P^l $ we can decompose it according
to (\ref{roz}), so that
$ p= \sum_{k=0}^{[ l/2]} r^{2k} h_k,\ h_k \in H^{l-2k}$.
It follows that
\begin{eqnarray}
\label{Roz}
\CO_q (p) = \sum_{k=0}^{[l/2]} \CR_q^{k}
\CO_q ( h_k ).
\end{eqnarray}
By virtue of the intertwining property (\ref{diag}) and (\ref{L_q})
we obtain: 
\begin{theo}\label{Weylharm}
The image of the space $H$ of harmonic polynomials in $P$ 
under the $q$-ordering map $\CO_q$ is independent of $q$. 
For $ w\in \CW$ we have 
\[
 w\in \CO_q (H ) \iff 
\sum_{j=1}^d [\, a_j,[\, a_j^+,w]]=0.
\]
\end{theo}
We shall denote this space by $\CH = \CO_q (H )$ and 
shall refer to it as the space of Weyl algebra harmonics.
We note that Geller in \cite{G2} observed, with a very different method 
and for the particular cases $q = 0, \pol, 1$ only, this somewhat surprising 
fact that the space of Weyl algebra harmonics does not depend 
on the ordering.

Getting back to (\ref{Roz}), we see that two factors are 
essential for understanding the structure of the Weyl algebra:
the space $\CH \subset \CW$ of Weyl harmonics and the action of 
operators $\CR_q$ on $\CW$. In fact, the equality (\ref{Roz}) 
leads to the following result.

\begin{theo}
Recall the space of radial polynomials in $P$, $J = \NC[r^2]$, 
and for $p(r^2) = \sum_{k=0}^l c_k r^{2k}\in J$ 
set $p(\CR_q) = \sum_{k=0}^l c_k \CR_q^k$.
Then the map
$$
J \times H \ni (p,h) \longrightarrow p(\CR_q) \CO_q(h) \in \CW
$$
gives rise to an isomorphism of $J \otimes H$ with $\CW$.
\end{theo}
In the sequel we shall investigate in detail the image 
$ \CJ_q = \CO_q(J)= \{ p(\CR_q)I \mid  p \in J \}$
of the space $J$ of invariant polynomials under the ordering map.
Although apparently depending on $q$ through its construction, it will turn 
out to be the same for all values of $q$ and consists of polynomials of a 
single element (the number operator (\ref{number-op})) of $\CW$ --- 
cf. Proposition \ref{indep} below.

For each nonnegative integer $k$ we set
\begin{equation}
\label{eta}
\eta_k = \eta_k [q,\,d] = \CR_q^k I,
\end{equation}
so that $\{ \eta_k \}$ is a natural basis of $\CJ_q$.
Let us write down the explicit form of a few first elements of this sequence.
Clearly $\eta_0=I$, and after collecting terms in (\ref{*}) and
making use of the CCR (\ref{CCR}) we arrive at
$$
\eta_1  = \sum_{j=1}^d a_j^+ a_j  +d(1-q)I = N + d(1-q)I.
$$
Here we have denoted by
\be \label{number-op}
 N = \sum_{j=1}^d a_j^+ a_j
\ee
an element  of the Weyl algebra, in physics called the number operator,
which plays a rather special role in questions related to the structure of
the Weyl algebra, see e.g. \cite{HO,me4,me5}. Its relevance for the present
situation stems from the fact that the space $\CJ_q$ consists of polynomials
of $N$. Before proving this fact, we bring in for illustration the
expression of  $\eta_2$ in terms of $N$.
Setting for simplicity $t_0 = d(1-q)$, one reads it as follows
\[
\eta_2 (N) =
(N + t_0 I)^2 + (1-2q) ( N + t_0 I) + q t_0 I.
\]
\begin{lemma} \label{omega}
For each $k \in \NZ_+$ there exists a unique
polynomial $\omega_k(t) = \omega_k (q,d;t)$ of precise degree $k$,
$\omega_k \in \NC [t] $,  such that
$$
 \eta_k [q,\,d]  =  \omega_k  (N).
$$
\end{lemma}
\Proof\ Clearly $\omega_0(t)\equiv 1$ and the form of
$\omega_1(t)$ and $\omega_2(t)$ can be read of the above formulae. To
prove the result in general we proceed by induction with respect to $k$.
First note the general relations (the so called pull-through relations)
\begin{equation} \label{pt}
a_i p (N) = p (N + 1) a_i, \qquad a_i^+ p (N) = p (N-1) a_i^+
\end{equation}
valid for any polynomial $p(N)$ of the
number operator. It follows that
\begin{equation}
\qquad \CR_q p ( N ) =  q(1-q)
(2N + d)  p (N) + (1-q)^2 (N + d) p (N+1) + q^2 N p (N-1).
\end{equation}
A comparison of highest order coefficients in terms at the right hand side
shows that applying $\CR_q$  to a polynomial in $N$ 
increases its degree by one. \hfill $\Box$

For completness we explicitly state the following
\begin{theo}\label{indep}
The space $\CJ_q$ is independent of $q$ and consists of 
polynomials in terms of the number operator $N$. Thus we set
\be
\CJ=\CJ_q=\NC[N].
\ee
\end{theo}

To determine the form of polynomials $ \omega_k$ we shall 
investigate the difference  and recurrence equations satisfied 
by them. We begin with the following observation.
\begin{lemma}
The space $\CJ$ is invariant under the action of the triple
$(\CR_q,\,\CL_q,\,\CE_q)$ and in fact $\eta_k$ given by {\rm(\ref{eta})}, with $k$
ranging over all nonnegative integers, form a weight basis for the ${\fr sl}_2$
action on $\CJ$ defined by it. Setting for convenience
$\eta_{-1} = 0$ this action is expressed by the well-known formulae
\begin{eqnarray*}
\CR_q    \eta_k  &\eq & \eta_{k+1}, \\
\CL_q    \eta_k &\eq & k(k+ d -1) \eta_{k-1}, \\
\CE_q    \eta_k &\eq & (2k+ d) \eta_k.
\end{eqnarray*}
\end{lemma}
This immediately follows by comparing the definition (\ref{eta}) and the
corresponding action of $\{R,\,L, \,E \}$ on the space of radial
polynomials.

By virtue of Lemma \ref{omega} we can transfer the action of
$\{\CR_q,\,\CL_q,\,\CE_q \}$ to the space of polynomials $\NC[t]$.
If $\CX$ is any from the triple of operators 
$\{\CR_q,\,\CL_q,\,\CE_q \}$ we define
the map $ \widetilde{X} : \NC[t] \mapsto \NC[t]$ by setting
\begin{eqnarray}
\label{tylda}
 ( \widetilde{X} p) (N)  = \CX p(N).
\end{eqnarray}
By the proof of Lemma \ref{omega}, the right hand side is a uniquely determined
polynomial of the number operator $N$.

Employing the classical notation for the forward and backward
difference operators, i.e. setting 
\begin{eqnarray*}
 \Delta \omega (t) = \omega (t+1) - \omega (t)
\qquad  \nabla \omega (t) = \omega (t) - \omega (t-1),
\end{eqnarray*}
we can state the following.
\begin{lemma}\label{difference-sl2}
The action of the triple $\{\CR_q,\,\CL_q,\,\CE_q \}$ on $\CJ$ 
induces the following action on $\NC[t]$.
\begin{eqnarray*}
\widetilde{R_q}   \omega ( t ) &\eq & 
(1-q)^2 (t + d)\Delta \omega (t) - q^2 t\nabla \omega(t) 
+ (t + d(1-q))\omega (t) \\
\widetilde{L_q} \omega ( t ) &\eq &
 (t + d)\Delta \omega(t) - t \nabla \omega(t),  \\
\widetilde{E_q} \omega ( t ) &\eq & 2(1-q)(t+d)
\Delta \omega (t) +
2qt\nabla \omega (t) +
d \omega(t).
\end{eqnarray*}
\end{lemma}
The proof is a routine calculation based on Lemma \ref{triple}  and (\ref{pt}).

A consequence of the relation
\[
\CE_q \eta_k  = \widetilde{E_q} \omega_k (N)= k\omega(N)
\]
is a difference equation satisfied by $\omega_k$ which  reads
\begin{eqnarray}
\label{Dif}
qt  \nabla \omega_{k}(t)
+(1-q) (t+d)
 \Delta \omega_{k} (t) = k \omega_{k} (t).
\end{eqnarray}

In order to fix a multiplicative constant for $\omega_k$ which is not
determined by the equation (\ref{Dif}), we look for an additional relation.
This is supplied by taking a suitable combination of operators appearing in
Lemma  \ref{difference-sl2}. One notes that the operator
$\widetilde{R_q} - q(1-q) \widetilde{L_q} +(q-\frac{1}{2})\widetilde{E_q}$
acts on polynomials of $t$ as multiplication by  the
factor $(t+ \frac{d}{2})$. Thus we see that the polynomials $\omega_k$ satisfy
the recurrence formula:
\begin{equation}
\label{Rec}
\qquad \qquad
\omega_{k+1}(t)-q(1-q)k(k+d-1)
\omega_{k-1}(t)-[t+(1-q)d-(2q-1)k]\omega_{k}(t) = 0.
\end{equation}
 Setting in addition $\omega_{-1}= 0$, the formula (\ref{Rec}) makes sense for
all $k\in \NZ_+\ (=\{0,\,1,\,2,\,\ldots\})$ and with the initial condition
$\omega_0 = 1$ determines the sequence $\omega_k$ uniquely. Now we can
formulate our first main result.

\begin{Theo} \label{theo}
Let\ ${}_{2}F_1 (a,b,c;x)$ be the Gauss hypergeometric function.
The polynomials $\omega_k$ defined in the Lemma {\rm\ref{omega}} 
are given by the formula
\[
\omega_k (t) = (d)_k (1-q)^k\, {}_2 F_1(-t,\,-k,\,d;\,\frac{1}{1-q}),
\]
where $(d)_k = d(d+1)(d+2)\ldots(d+k-1)$  denotes the shifted factorial.
\end{Theo}
\Proof
Let us first recall the well known Gauss recurrence formulae for
the hypergeometric function (cf. \cite{AA})
$$
(2b-c-bx+ax){}_2 F_1(a,\,b,\,c;\,x)
 + (c-b){}_2 F_1(a,\,b-1,\,c;\,x)
+b(x-1){}_2
F_1(a,\,b+1,\,c;\,x)=0.
$$
After the substitution of parameters
\[
a=-t, \qquad b=-k, \qquad c=d, \qquad x = \frac{1}{1-q},
\]
it coincides with the recurrence equation (\ref{Rec}).  
Since by definition $\omega_0=1$ and similarly ${}_2 F_1 (-t,\,0,\,d;\,\frac{1}{1-q})=1$
 we obtain the required result.
\hfill $\Box$
\medskip

\noindent{\bf Remarks.}
\medskip

1) For $q=1$ the expression for $\omega_k$ given in Theorem 1 should be
understood as a limit
 $$
\omega_k(t) = (d)_k \, \lim_{q \mapsto 1^-} (1-q)^k
{}_2 F_1(-t,\,-k,\,d;\,\frac{1}{1-q}).
$$

2) Other contiguous relations for ${}_2 F_1$ give other formulae fulfilled by
$\omega_k$. For example from the equation
$$
(2a-c-ax+bx){}_2 F_1 (a,\,b,\,c;\,x)+(c-a){}_2
F_1(a-1,\,b,\,c;\,x)+a(x-1){}_2 F_1(a+1,\,b,\,c;\,x)=0
$$
we obtain the difference equation (\ref{Dif}).

With a suitable change of variable it is possible to
obtain an explicit expression for the generating function for this  family of
polynomials. Recall that given a sequence of polynomials
$(g_k(\lambda))_{k=0}^{\infty}$, the function  $$
G(\lambda,s) = \sum_{k=0}^{\infty}g_k(\lambda) s^k
$$
is said to be a generating function for $(g_k(\lambda))_{k=0}^{\infty}$.

Now make the substitution
$ t+ t_0 = -\frac{i}{\alpha} \lambda$,
where $\alpha= (q(1-q))^{-1/2}$ and $t_0=(1-q)d$
and suitably renormalize the sequence of
polynomials by setting
\begin{eqnarray}
\label{gk}
g_k (\lambda) = g_k (q,d;\lambda)= \frac{i^k \alpha^k}{k!} \omega_{k}(t),
\quad {\rm where }
\quad \lambda =i \alpha (t+t_0).
\end{eqnarray}
Setting in addition $s_0 = i \alpha(q-\frac{1}{2})$,
we can rewrite the recurrence equation (\ref{Rec}) in the form
\begin{eqnarray}
\label{fg}
(k+2)g_{k+2}(\lambda) + (k+d)g_{k}(\lambda) + [-\lambda +
2s_0(k+1)]g_{k+1}(\lambda) = 0.
\end{eqnarray}
Multiplying both sides of (\ref{fg}) by $s^{k+1}$ and taking the sum over all
nonnegative $k$, we obtain an equation satisfied by the  generating function
for $(g_k(\lambda))_{k=0}^{\infty}$:
$$
(1 + 2s_0 s + s^2) \frac{\partial G}{\partial s} (\lambda,s) = (- d s +
\lambda) G (\lambda,s). $$
Performing integration over $s$, we arrive at the following result.

\begin{Theo}
Given $q \in [0,1]$ and $d \in \NN$  the function
\begin{eqnarray}
\label{G}
G(q,d;\lambda,s) = \frac{\exp{[\frac{2}{\alpha}(\lambda +d s_0) \arctan{(
\frac{2}{\alpha} (s+s_0) ))}}}
{[(s+s_0)^2 + \frac{\alpha^2}{4}]^{\frac{d}{2}}}
\end{eqnarray}
has the following expansion
$$
G(q,d;\lambda,s) = \sum_{k=0}^{\infty}g_k(\lambda) s^k.
$$
\end{Theo}

Now we we discuss in more detail the classical case $q=\pol$ of the Weyl
ordering. By a straightforward inductive argument we obtain
\begin{lemma} The polynowials $\omega_k(\pol,d;t)$ are even for even $k$, and
odd for odd $k$.
\end{lemma}
As the consequence, after the substitution (\ref{gk}), the polynomials
$g_k(\lambda)$ are real valued as functions of $\lambda = i \alpha (t+t_0)$.
Moreover, the previous results simplify to the following form.
\begin{coro}
The sequence  $(g_k)$ satisfies the recurrence equation
\begin{eqnarray}
 \label{rec}
(k+2)  g_{k+2} (\lambda) = \lambda g_{k+1} (\lambda ) - (k + d ) g_{k}
(\lambda),
\end{eqnarray}
and its generating function $G(d;\lambda,s) = G(\pol,d;\lambda,s)$ is given by
\begin{eqnarray}
\label{g1}
G(d;\lambda,s) =  \frac{ e^{\lambda
\arctan s}}{ ( \sqrt{ s^2 + 1} )^d}, \quad k=0,1,\ldots
\end{eqnarray}
\end{coro}

As we mentioned before, a special case (with $d=1$ and $q=\pol$)
of these results was obtained previously by C.~M.~Bender et al.
in \cite{Be2}, and more systematically by T.~H. Koornwinder in \cite{KO}.
In particular the connection of $g_k$ with the
continuous Hahn polynomials is observed there, and in the first
mentioned paper also the formula for the generating function
identical with the case of $d=1$ of our formula (\ref{g1}) was given.
However, neither the case of $q=\pol$ and $d>1$ nor other cases
of our results were mentioned there.

The next result identifies the polynomials $g_k(\lambda)$ with known
classes of orthogonal polynomials.

Recall (see \cite{KO}) that the continuous Hahn polynomials $p_k(x;a,b,c,d)$
are defined by $$
p_k(x;a,b,c,d)=i^k \frac{(a+c)_k(a+d)_k}{k!}
{}_3F_2(-k,k+a+b+c+d-1,a+ix;a+c,a+d;1)
$$
where ${}_3 F_2$ is generalized hypergeometric function (see \cite{AA}),
while the Meixner-Pollaczek polynomials $P^{(a)}_n(x;\phi)$ are given by
$$
P^{(a)}_n(x;\phi) = e^{in\phi } {}_2F_1 (-n,a+ix,2a;1-e^{-2i\phi}).
$$
The proof of the next result is a matter of standard calculation, when one 
takes into account the following formula connecting hypergeometric
functions, quoted in \cite{KO}.
\[
{}_2F_1 (-n,2a+2ix,4a;2) = {}_3F_2 (-n,n+4a,a+ix;2a,2a+\pol;1).
\]
\begin{Theo}
For  an arbitrary $d \in \NN$ and for $q=\pol$ the polynomials
$g_k(\lambda)=g_k(\pol,d;\lambda)$ given by {\rm (\ref{gk})} can be identified
as follows
\[
g_k(\lambda) =
 \frac{(d)_k}{(\frac{d}{2})_k (\frac{d}{2} + \pol
)_k} p_k \left(
 \frac{\lambda}{4};\frac{d}{4},\frac{d}{4} + \pol, \frac{d}{4},
\frac{d}{4} + \pol \right)
\]
and also
$$
g_k (\lambda) = \frac{(d)_k}{k!}
P^{(\frac{d}{2})}_k
\left( \frac{\lambda}{2};\frac{\pi}{2} \right).
$$
Moreover, they are orthogonal on $(-\infty,\infty)$ with respect to the
measure $\rho({\lambda})d\lambda$, where $\rho(\lambda) = |\Gamma
(\frac{d}{2} + i \frac{\lambda}{2} )|^2$.
\end{Theo}
Let us point out that the polynomials $\omega_k(t)$ do not share
certain properties of $g_k(\lambda)$. In particular by rewriting
the recurrence equation (\ref{Rec}) in the form
$$
\omega_{k+1}(t) = (A_k t + B_k) \omega_k (t) - C_k \omega_{k-1}(t)
$$
we see  that
$$
A_{k-1} A_k  C_k = -q(1-q)k(k+d-1) \leq
0.
$$
Thus the necessary condition for the existence of a positive
measure, with respect to which $\omega_k$ would be orthogonal,
is violated.

Also in the general case of nonsymmetric ordering corresponding to $q \not =
\pol$ the general picture is not so clear and satisfactory as in the
symmetric case.  From (\ref{Dif}) one could expect that the set
$\{\omega_k\}$ can be  described  as a subset of two known families of
polynomials --- Krawtchouk  ($K_n$) and Meixner ($M_n$).
In fact by comparing parameters we see that
\[
{}_2F_1(-t,\,-k,\,d;\,\frac{1}{1-q}) = K_k (t,\,1-q,\,-d) =
M_k(t,\,d,\,-\frac{1-q}{q}).
\]
However, there is again a problem with orthogonality,
since the parameters occuring in these formulae fall outside the range
required by the orthogonality condition (see \cite{AA} and \cite{NI}).

\section{Acknowledgements.} The ideas of this paper arose during
the doctoral studies of the first named author at the Department
of Mathematical Methods of Physics, Faculty of Physics,\
University of Warsaw under the supervision of the second named
author. Various parts of the results contained in this paper were
presented at the Banach Center ``Workshop on Lie groups and Lie
algebras'' in B\c{e}dlewo, September 2000 and Bia\l{}owie\.za
Workshop on Geometric Methods in Physics, July 2001. A fuller
presentation of the topic, discussing connections to other
classes of polynomials is contained in the doctoral dissertation
of the first named author.

Part of the work on this paper was done while the second named
author enjoyed the stay at the Humboldt University in Berlin
thanks to the support of SFB 288 {\sl Differential geometry and
quantum physics}.  This support and the hospitality of Professor
Dr. Thomas Friedrich are gratefully acknowledged.

After this work has been finished, Professor M. Engli\v{s} 
has pointed out to us the connection of our construction of the $q$-ordering
map $\CO_q$  with the extended symbol calculus discussed by A. Unterberger in 
{\it Encore des classes de symboles,\/} S\'eminaire Goulaouic-Schwartz 1977/1978, 
Expos\'e No. 6, \'Ecole Polytech., Palaiseau 1978. We intend to investigate this 
relation in more detail in a later work and we thank him for this indication.

Last but not least, we are grateful to the referee for his very careful reading of 
the manuscript and many remarks which led to an improvement of the paper.

\end{document}